\documentclass[runningheads]{llncs}
\usepackage[T1]{fontenc}
\usepackage{fontawesome}
\usepackage{graphicx}
\usepackage{color}
\usepackage{hyperref}
\hypersetup{hypertex=true,
            colorlinks=true,
            linkcolor=blue,
            anchorcolor=blue,
            citecolor=blue}
\usepackage{cite}
\usepackage{multirow}
\begin{document}
\title{Leukocyte Classification using Multimodal Architecture Enhanced by Knowledge Distillation}
\titlerunning{Leukocyte Classification using Multimodal Architecture}
\author{Litao Yang\inst{1}\textsuperscript{(\faEnvelopeO)}
\and Deval Mehta\inst{1} \and Dwarikanath Mahapatra\inst{2}\and Zongyuan Ge\inst{1}}
%index{Yang, Litao}
%index{Mehta, Deval}
%index{Mahapatra, Dwarikanath}
%index{Ge, Zongyuan}

\authorrunning{L. Yang et al.}
\institute{Monash Medical AI, Monash University, Melbourne, Australia \url{https://www.monash.edu/mmai-group} \\ \email{litao.yang@monash.edu} 
\and Inception Institute of Artificial Intelligence, Abu Dhabi, UAE}

\maketitle         % typeset the header of the contribution
\begin{abstract} Recently, a lot of automated white blood cells (WBC) or leukocyte classification techniques have been developed. However, all of these methods only utilize a single modality microscopic image i.e. either blood smear or fluorescence based, thus missing the potential of a better learning from multimodal images. In this work, we develop an efficient multimodal architecture based on a first of its kind multimodal WBC dataset for the task of WBC classification. Specifically, our proposed idea is developed in two steps - 1) First, we learn modality specific independent subnetworks inside a single network only; 2) We further enhance the learning capability of the independent subnetworks by distilling knowledge from high complexity independent teacher networks. With this, our proposed framework can achieve a high performance while maintaining low complexity for a multimodal dataset. Our unique contribution is two-fold - 1) We present a first of its kind multimodal WBC dataset for WBC classification; 2) We develop a high performing multimodal architecture which is also efficient and low in complexity at the same time.

\end{abstract}
\section{Introduction}
\begin{figure}[t!]
\includegraphics[width=0.5\linewidth]{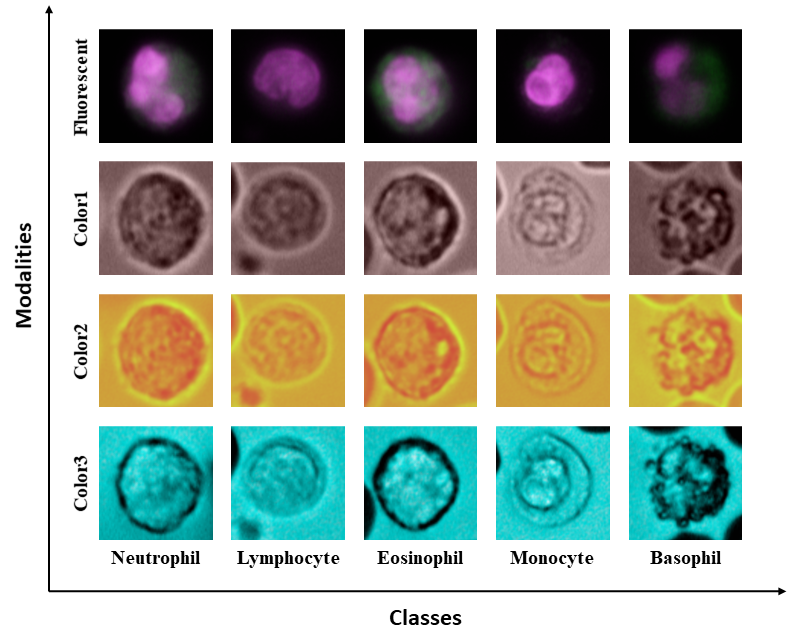}
\centering
\caption{WBC samples of our dataset which include four modalities and five classes} \label{fig1}
\end{figure}

A complete blood count (CBC) test is the foremost requirement for diagnosing any health-related condition of a person\cite{tkachuk2007wintrobe}, which consists of the count of red blood cells (RBCs), white blood cells (WBCs) and platelets\cite{theml2004color,george2003understanding}. Of these, WBCs are responsible to defend body organs and heal any damage to the biological structures\cite{blumenreich1990white}. Thus, it is vital for doctors to know the count of WBCs amongst the different categories to diagnose any specific disease or underlying health condition.
Recently, due to the advent of machine learning and deep learning, there have been a plethora of methods developed for automated detection and classification of WBCs\cite{khamael2020segmentation}. Most of these methods are based on blood smear microscopic images which provide rich features about the morphology and structure of the cells but require manual preparation and staining of the slides by trained personnel\cite{ramesh2012isolation}. More recently, the research community has developed image acquisition of cell images by mixing Acridine Orange (AO) dye with the blood samples and exposing them to different light source excitation yielding intense fluorescence like images\cite{das2021fluorescence}. The fluorescence based imaging is more efficient and can be integrated easily to some devices compared to the one based on blood smears\cite{forcucci2015all}. Some fluorescence based imaging methods have also been developed for automated detection and classification of WBCs\cite{yakimov2019label,das2021fluorescence}.
However, fluorescence based imaging suffers from phototoxicity and photobleaching, which makes the images less feature rich and the process of distinguishing between different cell types becomes more difficult as compared to the blood smear images\cite{ojaghi2020label}. Thus, we believe that multi-modal WBC images can be an effective solution for leveraging diverse features for making the models learn better and improve their classification performance\cite{baltruvsaitis2018multimodal}. To this end, we collect a unique WBC dataset with four light-sources - one from fluorescence and three from bright-field. Some samples from our dataset are shown in Fig.\ref{fig1}\footnote{\label{note1}Please note that the colors of the images are for representation only and not the actual colors. The technical details of excitation lights and actual colours will be released once the disclosure has been filed.}, where we represent the four modalities on the y-axis and the five WBC categories on the x-axis.

However, learning from a multi-modal dataset has its own challenges. One of the challenges is to devise a strategic fusion method that can leverage diverse features from the different modalities\cite{baltruvsaitis2018multimodal}. Based on the different possible locations of the fusion in an architecture, methods can be roughly divided into early and late fusion\cite{d2015review}. 
Early fusion\cite{snoek2005early} merges features from different modalities in the beginning and passes them to a single network to learn, whereas late fusion\cite{snoek2005early} employs separate networks for each modality and merges the features from each of them at the end. Both strategies have their pros and cons. While late fusion enables to increase the performance significantly by learning specific modality features using separate networks\cite{gadzicki2020early}, it also increases the overall framework complexity significantly. On the other hand, early fusion increases the framework complexity only slightly, however, it usually does not perform well as low-level fused features are not so effective for better learning by using a single network\cite{liu2016multispectral}. Apart from these two, there can be several different locations in the framework where fusion can be employed such as a halfway fusion~\cite{liu2016multispectral} or middle fusion\cite{damer2019learning}, but all these strategies will also have a trade-off between model complexity and performance.

In this work, for our multi-modal WBC dataset, we aim to address this trade-off of multimodal deep learning by developing low complexity networks which can achieve a higher performance. Specifically, we develop a framework for processing multimodal WBC images with tiny added cost of complexity compared with early fusion model while achieving an equivalent or even better performance than that of the late fusion strategy. We construct our framework in two steps (Fig.\ref{fig3}): First, 
we adapt the recently proposed Multi-input Multi-Output (MIMO)\cite{havasi2020training} architecture for our multi-modal dataset and develop a Multi-modal-MIMO (MM-MIMO) network for training modality specific independent subnetworks of a single deep learning network only. Second, on top of the MM-MIMO architecture, we further enhance its capability by employing knowledge distillation\cite{hinton2015distilling} to transfer the knowledge from late fusion strategy to MM-MIMO which helps to produce even better performance gains. We conduct extensive experiments to demonstrate the effectiveness of each part of our framework. To the best of our knowledge, we are the first to use a multi-modal dataset for WBC classification task, the first to adopt MIMO for multi-modal dataset, and the first to integrate knowledge distillation between different multimodal fusion methods. We believe our proposed idea can thus help to ease the trade-off between complexity and performance for multimodal datasets in general.
\section{Method}
\subsection{Sample Preparation and Data Collection}
We first collect blood samples from normal persons which are then exposed to the three different bright-field light sources and the resulting three different bright-field images are captured by a camera module. Then, we immediately stain the blood samples with Acridine Orange (AO) dye and expose them to another light source before collecting the resulting fluorescent images. Thus, we capture four different modal images of the blood samples (shown in Fig.\ref{fig1}), which are then given to two trained pathologist experts who then annotate and categorize the WBC cells images and build our multimodal WBC dataset.
\begin{figure}[t!]
\includegraphics[width=1\linewidth]{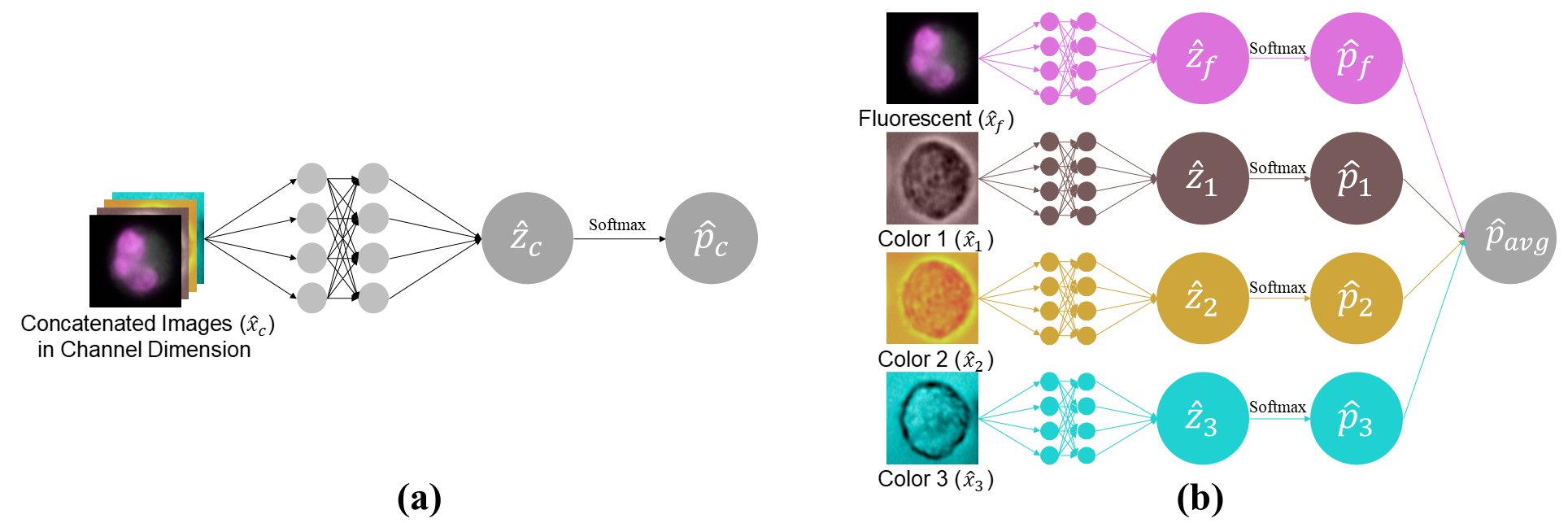}
\centering
\caption{Baseline fusion methods for our multimodal dataset (a): Early fusion; (b): Late fusion} \label{fig2}
\end{figure}
\begin{figure}[t!]
\includegraphics[width=1\linewidth]{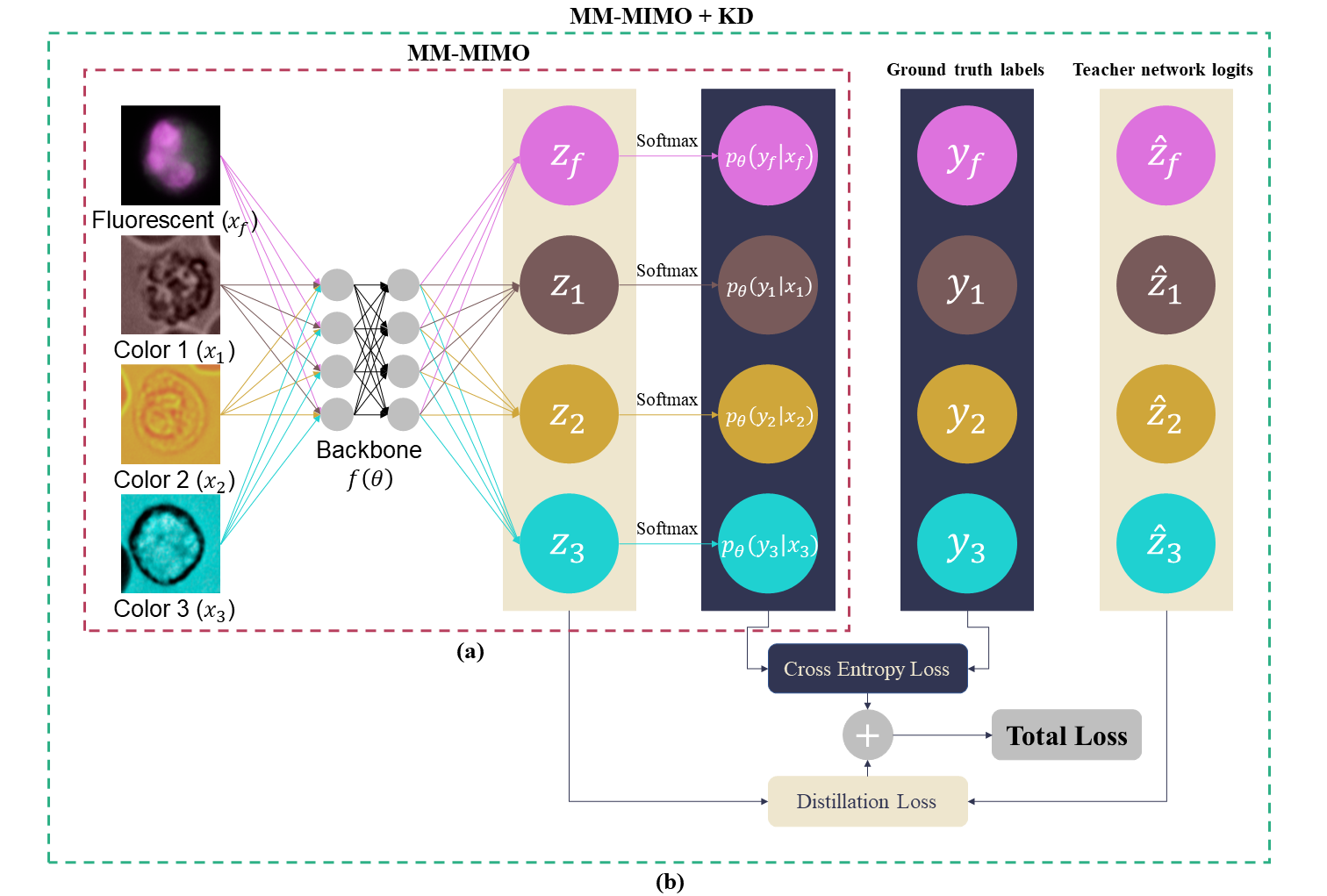}
\centering
\caption{The overview of our proposed framework. (a): Multi-modal Multi-input Multi-output Network architecture (MM-MIMO). We only use the cross entropy loss for training this part. (b): Multi-modal Knowledge Distillation (MM-MIMO + KD). We use combination of distillation loss and cross entropy loss for training this part.} \label{fig3}
\end{figure}
\subsection{Baseline Methods}
For our multimodal dataset, we first train two baseline methods of early fusion and late fusion as shown in Fig.\ref{fig2} (a) and (b). In early fusion, the four modal specific images from same WBC are concatenated in the channel dimension to \begin{math}\widehat{x}_{c}\end{math} which then passes through a single network to get the logits output \begin{math}\widehat{z}_{c}\end{math} and the probability prediction \begin{math}\widehat{p}_{c}\end{math}. In late fusion, we train four single networks independently based on each modality then use the same WBC with four modalities as input during forward pass and average their probability prediction \begin{math}\{\widehat{p}_{f},\widehat{p}_{1}, \widehat{p}_{2}, \widehat{p}_{3}\}\end{math} to get the final fusion prediction \begin{math}\widehat{p}_{avg}\end{math}. All of the four single networks and early fusion are trained by using the the standard cross entropy loss \begin{math}\mathcal{L}_{CE}\end{math}.
\subsection{Multi-modal Multi-input Multi-output Network (MM-MIMO)} 
Inspired by the sparse utilization of network parameters, a recently proposed architecture trained multiple subnetworks (MIMO) inside a single network for robust prediction of single modality datasets~\cite{havasi2020training}. We believe that MIMO can thus be used to learn independent modality specific subnetworks and hence we adapt it to suit it for our multi-modal dataset and develop a Multi-modal MIMO (MM-MIMO). Fig.\ref{fig3} (a) shows the architecture of MM-MIMO. Our proposed architecture of MM-MIMO treats the independent subnetworks as the separate individual modalities networks similar to that of a late fusion and predicts the classification category in a similar way to that of late fusion. However, MM-MIMO itself only consists of a single network while late fusion strategy consists of $N$ ($N$ being 4 for our case) different separate networks thus having $N$ times more complexity compared to MM-MIMO.

In the training process of MM-MIMO, the modal specific WBC images are concatenated in the channel dimension similar to that of early fusion. However, unlike early fusion, we randomly sample WBCs independently from the four modalities in MM-MIMO instead of using the same WBC as input. In other words, the input of MM-MIMO \begin{math}\{x_{f},x_{1}, x_{2}, x_{3}\}\end{math} belong to different WBC categories and their four different corresponding WBC labels \begin{math}\{y_{f},y_{1}, y_{2}, y_{3}\}\end{math} are used as the ground truth. The input data is then fed to a single backbone network \begin{math} f(\theta) \end{math} and corresponding four logit outputs \begin{math}\{z_{f}, z_{1}, z_{2}, z_{3}\}\end{math} are generated, which are then fed to a softmax layer to get the modal specific probability predictions \begin{math}\{p_{\theta}( y_{f} | x_{f}),p_{\theta}( y_{1} | x_{1}), p_{\theta}( y_{2} | x_{2}), p_{\theta}( y_{3} | x_{3})\}\end{math}. The overall loss \begin{math}L_{M}\end{math} shown in equation \ref{eq1} is the sum of the standard cross entropy loss \begin{math}\mathcal{L}_{CE}\end{math} for each of the four different modalities, which makes the whole network learn to classify the four different WBCs simultaneously and creating independent subnetworks within MM-MIMO. Since the input-output pairs are independent, the features derived from other modalities are not useful for predicting the corresponding modality output. During backpropagation, the corresponding subnetworks will learn to ignore the other modalities inputs and each output will consider and predict only the corresponding modality input. This way we get the independent subnetworks for each modality trained. During the testing phase, the four modality input images are from the same WBC category and the output predictions will be averaged to get the final prediction of that single WBC.
\begin{equation}\label{eq1}
L_{M}=\sum _{i=f,1,2,3}\mathcal{L}_{CE}\left( y_{i}, p_{\theta}\left(y_{i}| x_{i}\right)\right)
\end{equation}
\subsection{Multi-modal Knowledge Distillation}
Knowledge distillation (KD) has been widely used to transfer knowledge from teacher models to student models in order to improve the performance of student models\cite{gou2021knowledge}. Hinton et al.\cite{hinton2015distilling} first used the class probability derived from the teacher model as a soft label to guide the training of students. The teacher models are usually a high complexity network with higher performance than the compact student models. To further enhance the modality specific learning capability of MM-MIMO, we propose to utilize KD to transfer the knowledge from late fusion (acts as a high complexity teacher network) to MM-MIMO network (acts as a low complexity student network). Our inspiration comes from the idea that late fusion has separate powerful networks tailored to modality-specific features for the individual modalities, which have learnt the specific modality features better than those of low-complex individual subnetworks within MM-MIMO.
We show the framework for training MM-MIMO with KD in Fig.\ref{fig3} (b). Here our MM-MIMO model acts as the student network with logit outputs as \begin{math}\{z_{f}, z_{1}, z_{2}, z_{3}\}\end{math} which are to be guided from the teacher logit outputs \begin{math}\{\widehat{z}_{f},\widehat{z}_{1}, \widehat{z}_{2}, \widehat{z}_{3}\}\end{math}. These teacher logit outputs are produced from the four independent networks of the late fusion. During training, we enable independent subnetworks in MM-MIMO to learn from corresponding complex teachers in late fusion by employing KullbackLeibler divergence loss \begin{math}\mathcal{L}_{KL}\end{math} (given by equation~\ref{eq2}) between the student and teacher logits.
\begin{equation}\label{eq2}
L_{KD}=\sum _{i=f,1,2,3}\mathcal{L}_{KL}\left( p_{s}\left( \widehat{z}_{i},T\right) ,p_{s}\left( z_{i},T\right) \right)
\end{equation}
where \begin{math}p_{s}\left(z,T\right)\end{math} are soft targets \cite{hinton2015distilling} which are the probabilities of inputs belonging to the classes and contain the informative dark knowledge from the model. They can be computed from the  logits by a softmax function given by equation \ref{eq3}.
\begin{equation}\label{eq3}
p_{s}\left( z_{j},T\right) =\frac{{\exp \left( z_{j}/T\right) }}{\sum _{k}\exp \left( z_{k}/T\right)}
\end{equation}
where \begin{math}z_{j}\end{math} is the logit for the \begin{math}j\end{math}-th class in logits \begin{math}z\end{math} and \begin{math}T\end{math} is the temperature hyperaparameter. The total loss \begin{math}L_{T}\end{math} in multi-modal KD is the combination of distillation loss \begin{math}L_{KD}\end{math} and cross entropy loss \begin{math}L_{M}\end{math} in MM-MIMO, which is shown in equation \ref{eq4}. It is necessary to multiply distillation loss by \begin{math}T^2\end{math}, since when producing the soft targets with equation \ref{eq3}, the magnitudes of the gradients are scaled by \begin{math}\frac{1}{T^2}\end{math}. In this case, the relative contributions of hard and soft targets remain roughly the same and the independent subnetworks in MM-MIMO are learned under the supervision of both the ground truth and the cell-specific soft labels coming from the modality specific networks in late fusion framework.
\begin{equation}\label{eq4}
L_{T} =L_{M} + T^2L_{KD}
\end{equation}
\section{Experiment and Results}
\subsection{Dataset and Implementation}
In this study, we use the multi-modal WBC dataset collected by the method described in \textbf{2.1}. Our dataset consists of 14912 WBC samples with four modalities, which include 9616 Neutrophil (NEU), 4448 Lymphocyte (LYM), 677 Eosinophil (EOS), 124 Monocyte (MO), and 47 Basophil (BAS) in each modality.  

We perform five-fold cross-validation to evaluate our methods and report our results. We use the standard data augmentation - RandAugment\cite{cubuk2020randaugment} with N = 2 and M = 18 in all our experiments where N is the number of augmentation transformations to apply sequentially and M is the magnitude for all the transformations. Each WBC image is centercropped to a fixed size of 224×224. We \textbf{do not} use any other tricks for performance improvement in all experiments as we want to generalize our major contribution to the commonly employed strategies.
% \usepackage{multirow}
\begin{table}[t!]
\centering
\caption{Comparison of performance and network complexity of different methods.}
\label{tab1}
\resizebox{\textwidth}{!}{\begin{tabular}{|l|l|llll|ll|}
\hline
\multirow{2}{*}{Backbone} &
  \multirow{2}{*}{Method} &
  \multicolumn{4}{c|}{Performance Metrics} &
  \multicolumn{2}{c|}{Complexity} \\ \cline{3-8} 
 &
   &
  \multicolumn{1}{c|}{F1-score} &
  \multicolumn{1}{c|}{Sensitivity} &
  \multicolumn{1}{c|}{Specificity} &
  \multicolumn{1}{c|}{AUC} &
  \multicolumn{1}{c|}{FLOPs} &
  \multicolumn{1}{c|}{Params} \\ \hline
\multirow{9}{*}{Shufflenet V2} &
  Fluorescent &
  \multicolumn{1}{l|}{93.61 ± 0.97} &
  \multicolumn{1}{l|}{93.73 ± 0.94} &
  \multicolumn{1}{l|}{93.06 ± 1.47} &
  98.41 ± 0.26 &
  \multicolumn{1}{l|}{147.79M} &
  1.259M \\ \cline{2-8} 
 &
  Color1 &
  \multicolumn{1}{l|}{94.17 ± 0.55} &
  \multicolumn{1}{l|}{94.23 ± 0.61} &
  \multicolumn{1}{l|}{93.75 ± 0.98} &
  98.62 ± 0.31 &
  \multicolumn{1}{l|}{147.79M} &
  1.259M \\ \cline{2-8} 
 &
  Color2 &
  \multicolumn{1}{l|}{94.60 ± 0.42} &
  \multicolumn{1}{l|}{94.66 ± 0.45} &
  \multicolumn{1}{l|}{93.81 ± 1.35} &
  98.70 ± 0.32 &
  \multicolumn{1}{l|}{147.79M} &
  1.259M \\ \cline{2-8} 
 &
  Color3 &
  \multicolumn{1}{l|}{95.33 ± 0.70} &
  \multicolumn{1}{l|}{95.41 ± 0.65} &
  \multicolumn{1}{l|}{94.79 ± 1.32} &
  99.00 ± 0.44 &
  \multicolumn{1}{l|}{147.79M} &
  1.259M \\ \cline{2-8} 
 &
  Early fusion &
  \multicolumn{1}{l|}{95.43 ± 0.36} &
  \multicolumn{1}{l|}{95.52 ± 0.35} &
  \multicolumn{1}{l|}{94.91 ± 0.63} &
  99.04 ± 0.28 &
  \multicolumn{1}{l|}{172.17M} &
  1.261M \\ \cline{2-8} 
 &
  Late fusion &
  \multicolumn{1}{l|}{95.87 ± 0.31} &
  \multicolumn{1}{l|}{95.98 ± 0.35} &
  \multicolumn{1}{l|}{94.88 ± 1.05} &
  \textbf{99.43 ± 0.22} &
  \multicolumn{1}{l|}{591.16M} &
  5.036M \\ \cline{2-8} 
 &
  MM-MIMO &
  \multicolumn{1}{l|}{95.65 ± 0.46} &
  \multicolumn{1}{l|}{95.84 ± 0.44} &
  \multicolumn{1}{l|}{94.64 ± 0.93} &
  99.35 ± 0.12 &
  \multicolumn{1}{l|}{172.19M} &
  1.276M \\ \cline{2-8} 
 &
  Early fusion + KD &
  \multicolumn{1}{l|}{95.68 ± 0.49} &
  \multicolumn{1}{l|}{95.80 ± 0.46} &
  \multicolumn{1}{l|}{94.96 ± 0.85} &
  99.10 ± 0.40 &
  \multicolumn{1}{l|}{172.17M} &
  1.261M \\ \cline{2-8} 
 &
  \textbf{MM-MIMO + KD (Ours)} &
  \multicolumn{1}{l|}{\textbf{95.99 ± 0.52}} &
  \multicolumn{1}{l|}{\textbf{96.10 ± 0.50}} &
  \multicolumn{1}{l|}{\textbf{95.00 ± 0.95}} &
  99.32 ± 0.41 &
  \multicolumn{1}{l|}{\textbf{172.19M}} &
  \textbf{1.276M} \\ \hline
\multirow{9}{*}{Resnet34} &
  Fluorescent &
  \multicolumn{1}{l|}{93.67 ± 0.81} &
  \multicolumn{1}{l|}{93.78 ± 0.73} &
  \multicolumn{1}{l|}{93.14 ± 0.97} &
  98.24 ± 0.29 &
  \multicolumn{1}{l|}{3670.75M} &
  21.287M \\ \cline{2-8} 
 &
  Color1 &
  \multicolumn{1}{l|}{94.32 ± 0.90} &
  \multicolumn{1}{l|}{94.41 ± 0.87} &
  \multicolumn{1}{l|}{93.15 ± 1.72} &
  98.68 ± 0.20 &
  \multicolumn{1}{l|}{3670.75M} &
  21.287M \\ \cline{2-8} 
 &
  Color2 &
  \multicolumn{1}{l|}{94.82 ± 0.57} &
  \multicolumn{1}{l|}{94.87 ± 0.57} &
  \multicolumn{1}{l|}{94.43 ± 1.19} &
  98.76 ± 0.30 &
  \multicolumn{1}{l|}{3670.75M} &
  21.287M \\ \cline{2-8} 
 &
  Color3 &
  \multicolumn{1}{l|}{95.48 ± 0.52} &
  \multicolumn{1}{l|}{95.55 ± 0.47} &
  \multicolumn{1}{l|}{94.89 ± 1.23} &
  99.04 ± 0.22 &
  \multicolumn{1}{l|}{3670.75M} &
  21.287M \\ \cline{2-8} 
 &
  Early fusion &
  \multicolumn{1}{l|}{95.56 ± 0.63} &
  \multicolumn{1}{l|}{95.65 ± 0.58} &
  \multicolumn{1}{l|}{94.89 ± 1.05} &
  99.08 ± 0.48 &
  \multicolumn{1}{l|}{4024.79M} &
  21.315M \\ \cline{2-8} 
 &
  Late fusion &
  \multicolumn{1}{l|}{96.04 ± 0.48} &
  \multicolumn{1}{l|}{96.14 ± 0.43} &
  \multicolumn{1}{l|}{95.00 ± 1.02} &
  \textbf{99.44 ± 0.14} &
  \multicolumn{1}{l|}{14683.02M} &
  85.148M \\ \cline{2-8} 
 &
  MM-MIMO &
  \multicolumn{1}{l|}{95.81 ± 0.64} &
  \multicolumn{1}{l|}{95.94 ± 0.63} &
  \multicolumn{1}{l|}{94.76 ± 1.39} &
  99.28 ± 0.20 &
  \multicolumn{1}{l|}{4024.80M} &
  21.323M \\ \cline{2-8} 
 &
  Early fusion + KD &
  \multicolumn{1}{l|}{95.84 ± 0.51} &
  \multicolumn{1}{l|}{95.92 ± 0.49} &
  \multicolumn{1}{l|}{95.30 ± 0.63} &
  99.17 ± 0.23 &
  \multicolumn{1}{l|}{4024.79M} &
  21.315M \\ \cline{2-8} 
 &
  \textbf{MM-MIMO + KD (Ours)} &
  \multicolumn{1}{l|}{\textbf{96.13 ± 0.28}} &
  \multicolumn{1}{l|}{\textbf{96.24 ± 0.28}} &
  \multicolumn{1}{l|}{\textbf{95.33 ± 0.47}} &
  99.38 ± 0.13 &
  \multicolumn{1}{l|}{\textbf{4024.80M}} &
  \textbf{21.323M} \\ \hline
\end{tabular}}
\end{table}
\subsection{Quantitative Results}
In this section, we report the weighted F1-score, sensitivity, specificity and AUC for performance evaluation, and number of parameters and FLOPs for complexity evaluation. For comprehensive experimentation and to justify our proposed ideas, we experiment on two backbone architectures - an efficient backbone - ShuffleNet V2 and a massive backbone - Resnet34. We report all our experimentation results in Table \ref{tab1}.
\subsubsection{Baseline Performance}
In Table \ref{tab1}, we first note that for both the backbones, the performance of bright-field based networks (Color1, Color2, Color3) is superior compared to the fluorescence based. This is due to the fact that fluorescent image has relatively inferior features. Specifically, we note that Color3 performs the best amongst all the modalities with highest F1-score of \{95.33\%,95.48\%\} for Shufflenet V2 and Resnet34 respectively. Second, as expected, we note that late fusion increases the performance significantly compared to the early fusion. For late fusion, the f1-score is increased by \{0.44\%,0.56\%\} for Shufflenet V2 and Resnet34 compared to the best performing modality of Color3 respectively.
\subsubsection{Proposed Framework Performance}
For our proposed strategies, first we note that MM-MIMO increases the performance compared to the early fusion strategy, however, it cannot achieve a higher performance compared to late fusion. This is due to the fact that late fusion has highly complex independent networks for each modality whereas MM-MIMO has low complex independent subnetworks inside a single network. Thus, after integrating the KD with MM-MIMO, we can notice that the performance improves significantly which provides a comprehensive justification for utilizing KD to make the learning of the subnetworks within MM-MIMO stronger. Specifically, our overall strategy of MM-MIMO+KD achieves \{0.12\%,0.09\%\} higher f1-score than that of the high-performing late fusion strategy and surpasses it for both the backbones of Shufflenet V2 and Resnet34 respectively.

The results depicted in Table \ref{tab1} thus justify our proposed idea of combination of multiple modalities and also the specific advantages of each part of our overall proposed framework of MM-MIMO+KD for the task of WBC classification.
\subsubsection{Complexity Comparison}
It can be seen from the last column in Table \ref{tab1} that late fusion increases the model complexity significantly (four times in our case) compared to the early fusion approach. It should also be noted that our proposed approaches only increase the model complexity slightly. To be specific, MM-MIMO+KD only uses \{0.02M,0.01M\} more FLOPS compared to the early fusion strategy for Shufflenet V2 and Resnet34 respectively. This shows that our approach maintains a low complexity while achieving a high performance.
\begin{figure}[t!]
\includegraphics[width=1\linewidth]{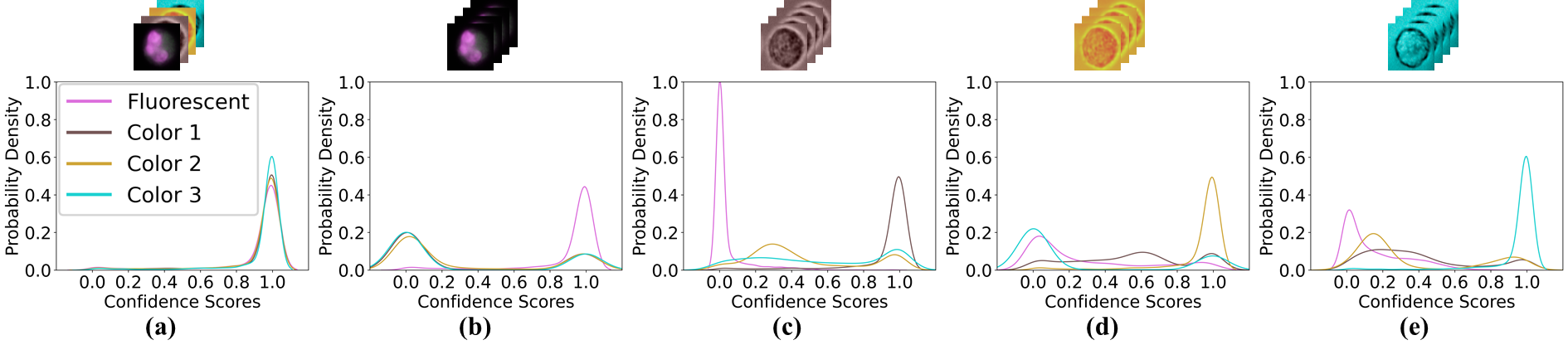}
\centering
\caption{Confidence score distributions of the four output heads in our MM-MIMO+KD framework with different input strategies (a): Four modalities; (b): Fluorescent only; (c): Color 1 only; (d): Color 2 only; (e): Color 3 only. (best viewed in zoom).} \label{fig4}
\end{figure}
\subsection{Visualization Results}
To better analyse the subnetworks independence within our MM-MIMO+KD architecture, we show a visualization of confidence scores of the modality-specific output heads in Fig.\ref{fig4}. We adopt five different input strategies for this visualization. First, all the four modalities are concatenated and given as the input to our framework. In this case, it can be seen that (Fig.\ref{fig4} (a)) all the four output heads are highly confident in their predictions as they receive their modality-specific input images. However, once we start to feed in only one specific modality image by concatenating it along the dimensions, only the corresponding output head of that modality gives a high confidence. The confidence score of the rest of the modality-specific heads reduces drastically as can be noted for each of the modality in Fig.\ref{fig4} (b) to (e). Thus, this visualization provides a clear understanding of the independence of our individual subnetworks within our framework.

\section{Conclusion}
In this work, we present a first of its kind leukocyte classification based on a multimodal dataset. Our extensive experimental results conclude that our developed multimodal framework enhanced with knowledge distillation has the capability of achieving a performance equivalent or even more than that of late fusion strategy while only slightly increasing the complexity parameters than that of a single network baseline. Thus, our proposed approach helps to ease the performance-complexity trade-off for a multi-modal dataset significantly. We believe that our proposed framework can act as a guidance to the research community working with multimodal data to help create efficient multimodal networks.

\bibliographystyle{splncs04}
%\cite{havasi2020training}
%\cite{bain2005diagnosis}
\bibliography{ref}
%





% LNCS Homepage, \url{http://www.springer.com/lncs}. Last accessed 4
% \end{thebibliography}
\end{document}